\documentclass[aps,prb,superscriptaddress,twocolumn,showpacs,amsmath,amssymb,floatfix]{revtex4}% Physical Review B
\usepackage{setspace}
\usepackage{array}
\usepackage{graphicx}% Include figure files
\usepackage{dcolumn}% Align table columns on decimal point
\usepackage{bm}% bold math
\usepackage[latin1]{inputenc}
\usepackage{graphicx,color,epsfig,rotate}
\usepackage{xspace}
\usepackage{hyperref}
\bibdata{references.bib}

\newcommand{\BaCo}{Ba(Fe$_{1-x}$Co$_{x}$)$_{2}$As$_{2}$\xspace}
\newcommand{\BaRu}{Ba(Fe$_{1-x}$Ru$_{x}$)$_{2}$As$_{2}$\xspace}
\newcommand{\BaK}{Ba$_{1-x}$K$_x$Fe$_{2}$As$_{2}$\xspace}
\newcommand{\BaKCo}{Ba$_{1-x}$K$_x$(Fe$_{1-y}$Co$_{y}$)$_{2}$As$_{2}$\xspace}
\newcommand{\BaKCos}{Ba$_{1-z}$K$_z$Fe$_{1.86}$Co$_{0.14}$As$_{2}$\xspace}

\newcommand{\BaFA}{BaFe$_{2}$As$_{2}$\xspace}

\newcommand{\BaP}{BaFe$_{2}$(As$_{1-x}$P$_x$)$_{2}$\xspace}
\newcommand{\mb}{M\"ossbauer\xspace}
\newcommand{\mbs}{M\"ossbauer spectroscopy\xspace}
\newcommand{\musr}{muon spin relaxation\xspace}

\pdfoutput=1
\begin{document}

\title{Microscopic Coexistence of Magnetism and Superconductivity in charge compensated {\BaKCo}}

%%%% AUTHORS %%%%%%%%%%%%%%%%%%%%%%%%%%%%%%%%%%%%%%%%%%%%%%
\author{Til Goltz}
\affiliation{Institute of Solid State Physics, TU Dresden, D--01069 Dresden, Germany}
\author{Veronika Zinth}
\affiliation{Department Chemie der Ludwig-Maximilians-Universit\"{a}t M\"{u}nchen, D--81377 M\"{u}nchen, Germany}
\author{Dirk Johrendt}
\affiliation{Department Chemie der Ludwig-Maximilians-Universit\"{a}t M\"{u}nchen, D--81377 M\"{u}nchen, Germany}
\author{Helge Rosner}
\affiliation{Max Planck Institute for Chemical Physics of Solids, D--01187 Dresden, Germany}
\author{Gwendolyne Pascua}
\affiliation{Laboratory for Muon--Spin Spectroscopy, Paul Scherrer Institut, CH--5232 Villigen PSI, Switzerland}
\author{Hubertus Luetkens}
\affiliation{Laboratory for Muon--Spin Spectroscopy, Paul Scherrer Institut, CH--5232 Villigen PSI, Switzerland}
\author{Philipp Materne}
\affiliation{Institute of Solid State Physics, TU Dresden, D--01069 Dresden, Germany}
\author{Hans-Henning Klauss}
\affiliation{Institute of Solid State Physics, TU Dresden, D--01069 Dresden, Germany}
%%%%%%%%%%%%%%%%%%%%%%%%%%%%%%%%%%%%%%%%%%%%%%%%%%

\date{\today}% It is always \today, today,
    % but any date may be explicitly specified

\begin{abstract}
We present a detailed investigation of the electronic phase diagram of effectively charge compensated \BaKCo with $x/2\approx y$. Our experimental study by means of x-ray diffraction, \mbs, \musr and ac~susceptibility measurements on polycrystalline samples is complemented by density functional electronic structure calculations.
For low substitution levels of $x/2\approx y \leq 0.13$, the system displays an orthorhombically distorted and antiferromagnetically ordered ground state. The low temperature structural and magnetic order parameters are successively reduced with increasing substitution level. We observe a linear relationship between the structural and the magnetic order parameter as a function of temperature and substitution level for $x/2\approx y \leq 0.13$.
At intermediate substitution levels in the range between 0.13 and 0.19, we find superconductivity with a maximum $T_c$ of 15\,K coexisting with static magnetic order on a microscopic length scale. % as we can prove for two samples with $x/2 \approx y = 0.16$ and $0.19$.
For higher substitution levels $x/2\approx y\geq0.25$ a tetragonal non-magnetic ground state is observed.
Our DFT calculations yield a significant reduction of the Fe 3d density of states at the Fermi energy and a strong suppression of the ordered magnetic moment in excellent agreement with experimental results.
The appearance of superconductivity within the antiferromagnetic state can by explained by the introduction of disorder due to non-magnetic impurities to a system with a constant charge carrier density.
\end{abstract}

% PACS, the Physics and Astronomy Classification Scheme.
%\keywords{Suggested keywords}%Use showkeys class option if keyword display desired

\maketitle
%
%
%%%%%%%% main part %%%%%%%%
\section{Introduction}
Shortly after the discovery of superconductivity in the electron-doped 1111 ferropnictides,\cite{hosono1111} the structural related $A$Fe$_{2}$As$_{2}$-based compounds ($A$=Ba, Sr and Ca) were also found to display superconductivity when magnetic order is sufficently suppressed by chemical substitution or applying pressure.\cite{RotterPhysRevLett.101.107006,Schnelle-Sr122-masterpaper-PhysRevLett.101.207004,Ca122masterpaper-Ronning-JPhsyCondMat-20-322201,122PressureMasterpaperJPhysCondMat-21-012208,ReviewPaglioneGreene,ReviewCanfieldBudko}

Changing the nominal electron count $e$ by introducing transition metal (TM) dopants on the Fe site or replacing the alkaline earth by an alkaline metal on the Ba site in \BaFA is one strategy. Electron doping on the Fe site using  TM = Co, Ni, Rh, Pd, Ir and Pt in Ba(Fe$_{1-x}$TM$_{x}$)$_{2}$As$_{2}$ leads to superconductivity and coincident phase diagrams for the 3d and 4d elements were found after appropriate scaling of $x$ and $e$.\cite{SefatPhysRevLett.101.117004,PhysRevB.80.024511,PhysRevB.82.024519,RhIrPd-doped-Ba122-PhysRevB.80.024506,Pt-doped-Ba122-JPhysConMat-22-072204} This includes the suppression of the structural (tetragonal-to-orthorhombic) and the antiferromagnetic (AFM) phase transition temperatures ($T_S$, $T_N$) and the appearance of a superconducting dome.
Hole doping on the Ba site by substitution with K likewise introduces superconductivity\cite{Rotter2008} and the obtained \textit{T}-\textit{e}~phase diagrams for \BaK and \BaCo show similar properties: $T_S$ and $T_N$ are suppressed with increasing substitution and superconductivity appears once the AFM order is sufficiently weakened. In addition, there is conclusive evidence for microscopic coexistence of magnetism and superconductivity prior to the full suppression of the magnetic order in both systems.\cite{Pratt-CoexistenceInCoDoped122-PhysRevLett.103.087001,Wiesenmayer2011}
In contrast to hole doping on the Ba site, in-plane substitution of Fe by Mn\cite{Mn-dopedBa122-Liu2010S513} and Cr\cite{PhysRevB.79.224524} does not introduce superconductivity. Also, electron doping with Cu does not support superconductivity \cite{ReviewCanfieldBudko} except below 2\,K in the vicinity of $x_{\text{Cu}}\approx0.044$.\cite{PhysRevB.82.024519}
However, the precise role that each substituent plays in controlling magnetism and/or the appearance of superconductivity is still under debate. In particular, the effect of impurity scattering, disorder and their impact on the electronic structure or the dichotomy between localized and itinerant physics are still matters of discussion.\cite{PhysRevB.82.024519,PhysRevLett.105.157004,berlijn-PRL108-207003,PhysRevLett.110.107007,PhysRevB.87.134516}

The study of systems with nominal isovalency as realized in the quaternary systems \BaRu\cite{Sharma-RuDopedMasterpaper-PhysRevB.81.174512} and \BaP\cite{Jiang-BaFeAsPmasterpaper-JPhysCondMat21-382203} is thus straightforward to get a better understanding of this issue. Even without carrier doping, both systems display superconductivity albeit much higher substitution levels of $x_{\text{Ru}}\geq 0.18$ and $x_{\text{P}}\geq0.2$ are needed to sufficiently supress $T_N$ and $T_S$. Notably, the case of Ru substitution is suitable for a comparision to transition metal substitution since the only modifications within the FeAs plane are done on the same site. A similar approach in which the in-plane modifications are limited to the substitution on the Fe site while maintaining nominal isovalency was proposed by Suzuki et. al.:\cite{Suzuki2010} They report superconductivity by combining Co and K substitution in the quintary system \BaKCo with $x/2\approx y=0.14-0.22$.
In the context of studying the influence of increasing in-plane substition level on structure, magnetism and superconductivity, the charge compensated \BaKCo system is of particulary interest, since the Co states strongly resemble the Fe 3d states and the impurity potential of Co ($U_\text{eff}=1.8\pm0.6$ eV) and Fe ($U_\text{eff}=1.4\pm0.6$ eV) is rather similar.\cite{PhysRevB.87.134516}

We synthesized and studied a series of polycrystalline samples of charge compensated \BaKCo with a substitution level $x/2\approx y$ ranging from 0 to 0.25. We complement our experimental results with a set of density functional theory calculations aiming to seperate the effects of the crystallographic changes and the influence of double substitution onto the electronic structure. This investigation of the electronic phase diagram also continues erlier work on the solid solution \BaKCos, where we have shown, that a combination of Co and K substitution leads to an effective compensation of hole and electron doping near nominal charge compensation $z=0.14$.\cite{Zinth2011} This result is in good agreement with photoemission data, showing a rigid-band-like filling or depleting of the electronic states near the Fermi level.\cite{PhysRevB.83.094522} The result, that electron (hole) Fermi surface volumes increases (decreases) with substitution, is qualitatively consistent with a rigid-band model for TM=Co.\cite{PhysRevLett.110.107007} We conclude as a basic premise for this work, that for nominal charge compensated \BaKCo, the concept of isovalency due to effective electron and hole compensation is valid.
\section{Methods}
Polycrystalline samples were prepared from stoichiometric mixtures of Ba, K and (Fe$_{1-y}$Co$_{y}$)$_{2}$As$_{2}$ in alumina crucibles, sealed in silica tubes and heated to 913\,K. The samples were annealed at 983\,K for 10\,h, homogenized to a fine powder and annealed at 1063\,K two times for 10\,h. The Ba:K ratio was determined by the Rietveld analysis yielding an uncertainity for $x/2\approx y$ of $\pm0.02$. Details of the synthesis method are described in Ref.~\cite{Zinth2011} Powder diffraction data were collected on a Huber G670 diffractometer with Co-K$_{\alpha1}$ or Cu-K$_{\alpha1}$-radiation, equipped with a closed cycle He cryostat.
From the AC susceptibility measurements, the superconducting transition temperature was extracted by fitting the steepest descent with a line and using the point of intersection with the normal-state susceptibility as $T_c$ (onset).
M\"ossbauer spectra were recorded in a standard transmission geometry setup using a $^{57}$Co/Rh source with an emission line witdh (FWHM) of $\Gamma=0.27(1)\,$mm/s. The spectrometer was calibrated to $\alpha$-Fe at room temperature. The powdered sample was mixed with methanol in a thin PA6.6 container to ensure a homogeneous surface thickness of 5 to 8\,mg\,Fe/cm$^2$. A CryoVac Konti IT cryostat with He exchange gas was used to stabilize temperatures between 4.2\,K and 300\,K.
Muon spin relaxation measurements were performed using the GPS spectrometer at the $\pi$M3 beamline of the Swiss Muon Source at the Paul Scherrer Institut, Switzerland. The data was analyzed with the MUSRFIT package.\cite{Suter2012}
Scalar-relativistic density functional (DFT) electronic structure calculations were performed using the full-potential FPLO code,\cite{DFT.PhysRevB.59.1743,DFT.PhysRevB.60.14035} version fplo9.01-35.\footnote{For version details see http://www.fplo.de} For the exchange-correlation potential within the local density approximation (LDA) the parametrizations of Perdew-Wang\cite{DFT.PhysRevB.45.13244} was chosen. The calculations were carried out on a well converged mesh of 1728 k-points in the Brillouin zone (12x12x12 mesh) to ensure a high accuracy for details in the electronic density. The partial Ba substitution with K and Fe with Co, respectively, was modeled within the virtual crystal approximation (VCA).\cite{DFT.VCA.PhysRevB.85.035207} For each substitution, the calculations were carried out for the experimental lattice parameters\cite{zinthdiss} if available, otherwise the structural data were linearly interpolated between neighboring experimental data points (see Fig.~\ref{fig:n43str}). For the simulation of the stripe antiferromagnetic order a $\sqrt{2}\times\sqrt{2}$ supercell within the tetragonal plane was chosen. The As-\textit{z} position was optimized with respect to the total energy.
\section{Results}
\subsection{Structure}
\begin{figure}[tbp]
\center{
{\includegraphics[height=100mm,angle=0,width=0.45\textwidth]{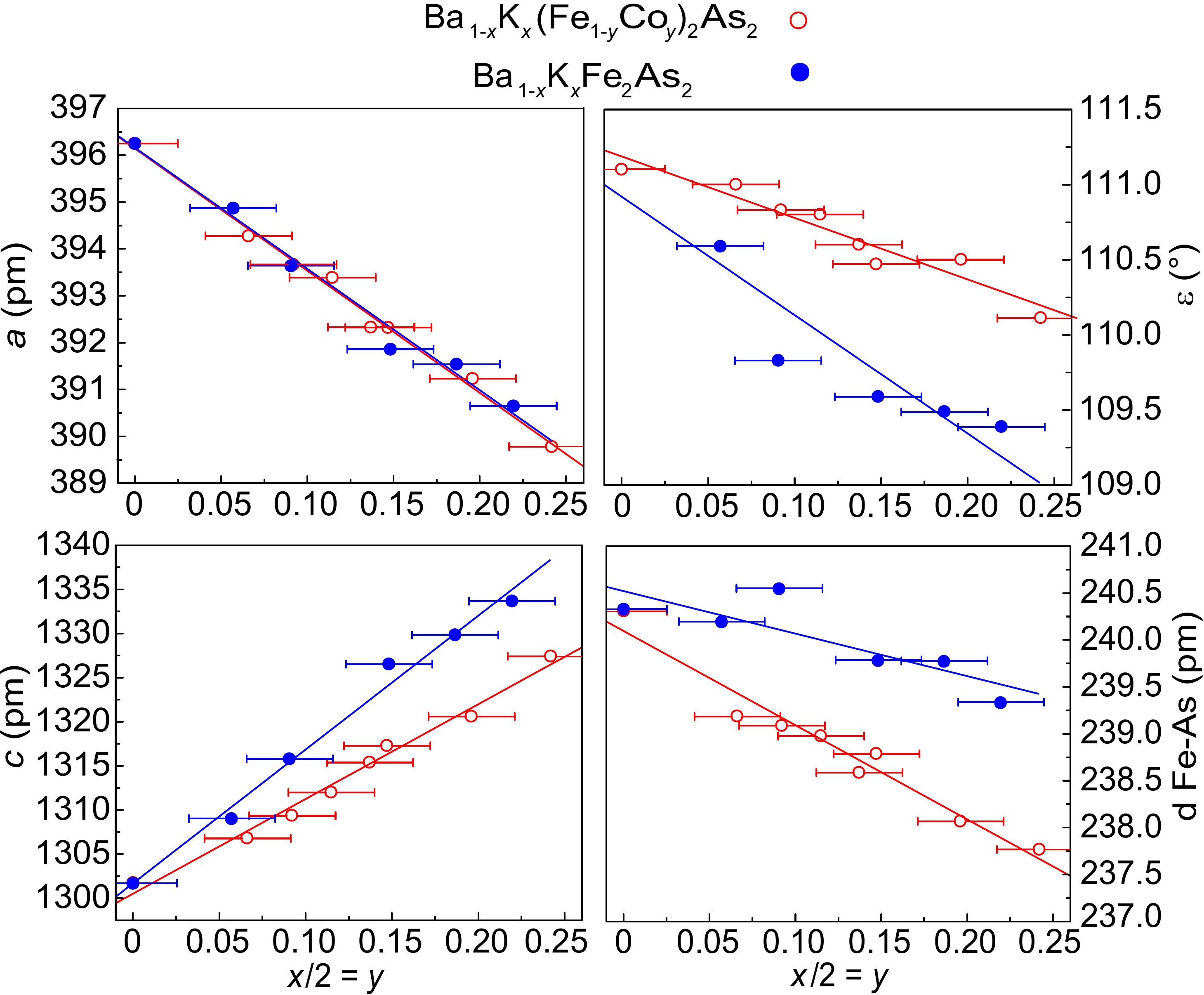}}
\caption{Structural changes in charge compensated {\BaKCo} (red, unfilled symbols) compared to {\BaK} (blue, filled symbols) with increasing potassium content per FeAs-layer ({\it x}/2). Lattice parameter {\it a} (top left) and {\it c} (bottom left), As-Fe-As angle $\varepsilon$ (top right) and Fe-As distance {\it d} (bottom right). %Note that {\it x}/2 $\approx$ {\it y}.
Data for {\BaK} is taken from Ref.~\cite{Rotter2008}}
\label{fig:n43str}
}
\end{figure}
In Fig.~\ref{fig:n43str}, we show the structural parameters of charge compensated {\BaKCo} (red) at room temperature compared to hole doped {\BaK}\cite{Rotter2008} (blue) to emphazise the changes due to the additional in-plane Co substitution. While {\it a} is not affected by the additional cobalt substitution and decreases at the same rate as in {\BaK}, the increase of {\it c} is clearly reduced in {\BaKCo}. Consequently, the value of the As-Fe-As angle $\varepsilon$ changes less for {\BaKCo} than reported for {\BaK} and reaches only $\sim$110.5$^\circ$ at {\it x}/2 = 0.2, where the ideal tetrahedral angle of 109.47$^\circ$ is found in {\BaK}. Interestingly, additional Co doping also leads to a decrease of the Fe-As distance; a reduction of $\sim$1\% is found for {\it x}/2 = 0.25 compared to only $\sim$0.3\% in {\BaK}.

\begin{figure}[tbp]
\center{
{\includegraphics[height=80mm,angle=270,width=0.48\textwidth]{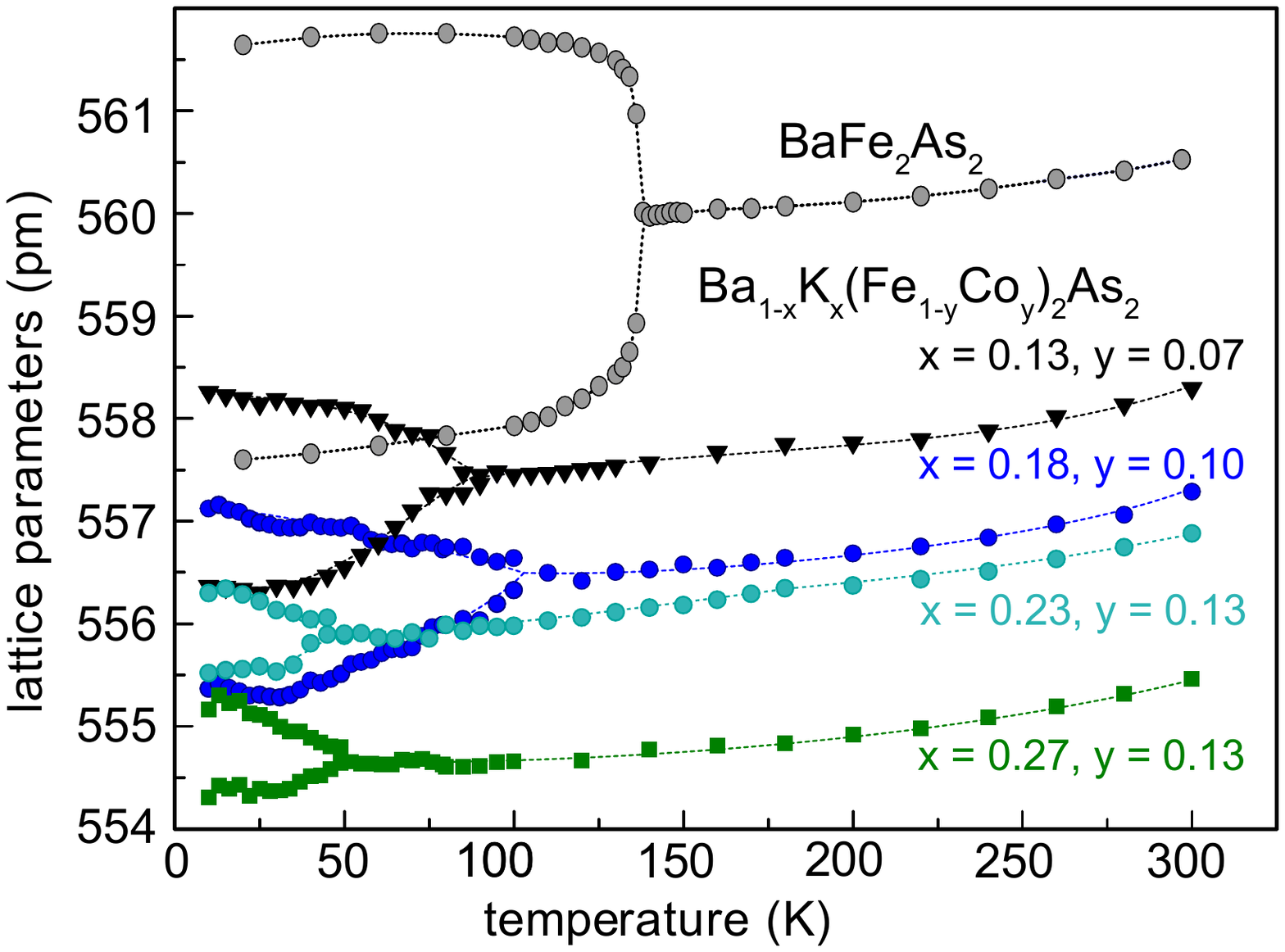} }
\caption{Lattice parameters of {\BaKCo} with {\it x}/2 $\approx$ {\it y} = 0.00 - 0.13 (in orthorhombic notation), showing the orthorhombic splitting at low temperatures. Data for {\BaFA} is taken from Ref.~\cite{Rotter2009}}
\label{fig:n43tt}
}
\end{figure}

As previously reported, \cite{Zinth2011} charge compensated {Ba$_{0.87}$K$_{0.13}$Fe$_{1.86}$Co$_{0.14}$As$_{2}$} shows the tetragonal to orthorhombic phase transition, but with a reduced orthorhombic distortion compared to {\BaFA}. To check how this parent like phase is influenced by increased cobalt and potassium substitution in the charge-compensated state, low temperature powder diffraction was performed. The results, depicted in Fig.~\ref{fig:n43tt}, show that the splitting of the lattice parameters becomes weaker with increasing substitution level. The splitting is significantly reduced from $\Delta$ = {\it b} $-$ {\it a} = 4\,pm in {\BaFA} to $\sim$ 2\,pm for {\it x}/2 $\approx$ {\it y} $\approx$ 0.07 and 0.09 and further to $\sim$\,0.8\,pm for {\it x}/2 $\approx$ {\it y} $\approx$ 0.13, where only a broadening of reflections at low temperatures is observed. Due to the diminutive size of the splitting it is difficult to determine the exact phase transition temperatures by the Rietveld method, but the general trend - a decrease of the structural phase transition temperature {\it T}$_s$ - is clearly visible. Thus we observe a decrease of both orthorhombic distortion and phase transition temperatures with increasing substitution level {\it x}/2 $\approx$ {\it y} in charge compensated {\BaKCo}, which means that the structural phase transition is subsequently suppressed.
%
% ----------------------------------------------------
%
\subsection{Magnetism}
The magnetic order was microscopically studied by $^{57}$Fe~\mbs and muon spin relaxation on identical samples of \BaKCo.

\begin{figure}
\centering
  \label{fig:mb01}\includegraphics[width=0.4\textwidth]{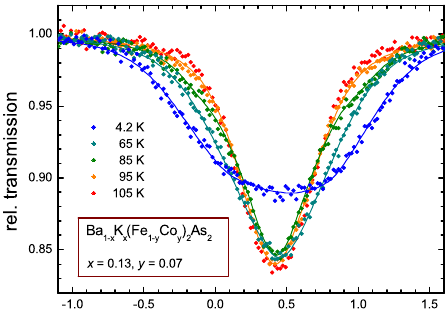}
  \label{fig:mb02}\includegraphics[width=0.4\textwidth]{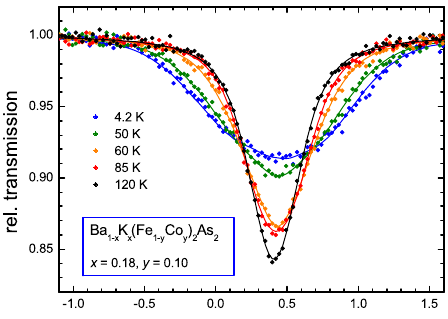}
  \label{fig:mb03}\includegraphics[width=0.4\textwidth]{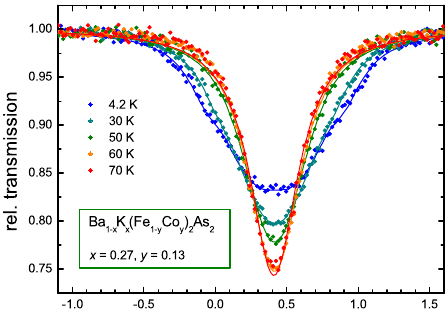}
  \label{fig:mb04}\includegraphics[width=0.4\textwidth]{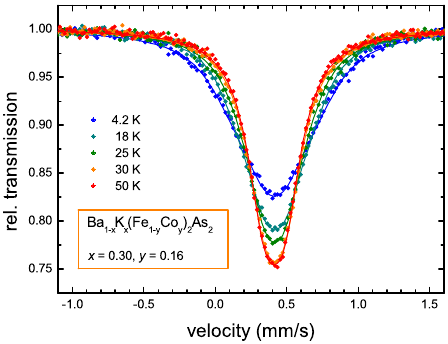}
  \vfill
\caption{Represantative $^{57}$Fe M\"ossbauer spectra of {\BaKCo} for {\it x}/2 $\approx$ {\it y} = 0.07, 0.10, 0.13 and 0.16. The magnetic phase transition is evidenced by the line broadening and the concomitant transfer of spectral intensity from the central resonance to the shoulders (see text for details).}
\label{fig:mbspectra}
\end{figure}

\begin{figure}
\centering
  \includegraphics{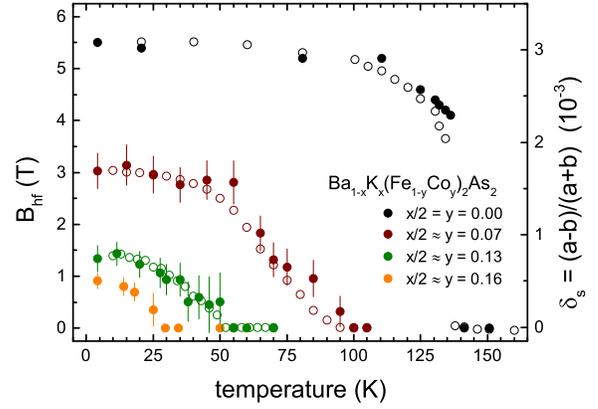} %[width=0.45\textwidth]
\caption{Temperature dependence of the weighted hyperfine field $B_{\text{hf}}$ (full circles, left scale) and the orthorhombic splitting parameter $\delta_s$ (open circles, right scale). Experimental data for \BaFA ({\it x}/2 = {\it y} = 0) is taken from Ref.~\cite{Rotter2009}}
\label{fig:structurvsmagnetism}
\end{figure}

Temperature dependent \mb~spectra were recorded and the temperature dependence of the iron hyperfine field $B_{\text{hf}}(T)$ was extracted. Since it is proportional to the on-site magnetic moment, it traces the magnetic order parameter for a later comparison to the structural order parameter defined by the normalized orthorhombic splitting. Representative low temperature spectra illustrating the magnetic phase transition are shown in Fig.~\ref{fig:mbspectra}.
No resolved lines from magnetic nuclear Zeeman splitting can be observed even at lowest temperatures. Due to the small ordered magnetic moment and the broad field distribution resulting from the high in-plane cobalt substitution level, the onset of magnetic order is deduced directly from the line broadening seen in the raw data: Upon decreasing the temperature below the onset of the antiferromagnetic ordering, a transfer of spectral weight from the center of the paramagnetic, singlet-like resonance to its shoulders is observed. ${T^{\text{onset}}_{M}}$ (index \textit{M} for \mbs) is accordingly defined by the highest temperature below which the intensity of the central resonance starts to decrease. Below ${T^{\text{onset}}_{M}}$, spectra were analysed with a Gaussian distribution of effectively static iron hyperfine fields.\cite{Rancourt1991} The error bars for the hyperfine fields given in Fig.~\ref{fig:structurvsmagnetism} represent the widths $\sigma_{\text{hf}}=\left(2 \ln 2 \right)^{-1/2}\times$ HWHM of the Gaussian field distribution. Two components were used to seperate magnetic ($\mathcal{M}$) and non-magnetic (1-$\mathcal{M}$) volume fractions: In the field distribution, the non-magnetic contribution is centered at zero with a fixed distribution width ($\left\langle{B_{\text{hf}}}\right\rangle=0$~T, $\sigma_{\text{hf}}=0.5$~T)\footnote{$\sigma_{\text{hf}}=\,0.5$ T was chosen, so that $2\sigma_{\text{hf}}=\Gamma/4$. This value corresponds to the experimental resolution limit due to the $^{57}$Co/Rh sources line width (FWHM) of $\Gamma=0.27$~mm/s.} whereas the magnetic contribution displays a non-zero hyperfine field ($\left\langle{B_{\text{hf}}}\right\rangle \geq0.5\,\text{T},~\sigma_{\text{hf}}$ free). The Lorentzian linewidth was fixed to its value from just above ${T^{\text{onset}}_{M}}$, since it is severe correlated to $\sigma_{\text{hf}}$. The magnetic volume fraction saturates at $100$\% for the purely magnetic samples with {\it x}/2 $\approx$ {\it y} = 0.07 and 0.10 and at $\sim45$\% for the partially superconducting samples with {\it x}/2 $\approx$ {\it y} = 0.13 and 0.16 at low temperatures.
%% definition of parameters
%
We define the antiferromagnetic transition temperature $T_N=T^{50\%}_{M}$ by the 50\% volume fraction criterium of the low temperature saturation value. The values for $T^{50\%}_{M}$ are obtained from sigmoidal fits of the temperature dependence of $\mathcal{M}(T)$ and given in Tab.~\ref{tab:moessparameter}.
The temperature dependence of the weighted hyperfine field $B_{\text{hf}}=\left\langle{B_{hf}}\right\rangle \times \mathcal{M}$ is shown together with the orthorhombic splitting parameter $\delta_S=\left(a-b\right)/\left(a+b\right)$ in Fig.~\ref{fig:structurvsmagnetism}.

\begin{figure}
	\centering
		\includegraphics{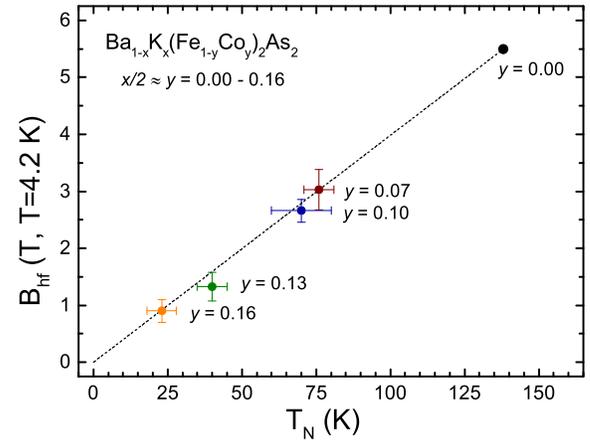} %[width=0.50\textwidth]
	\caption{The low temperature value of the weighted hyperfine field $B_{\text{hf}}$(T=4.2\,K) and $T_N$ show a linear relationship for all magnetic samples. The experimental data for \BaFA ($y=0$) is taken from Ref.~\cite{Rotter2009}}
	\label{fig:bhf-vs-tn-graph}
\end{figure}

For all compositions, the curves of the structural and the magnetic order parameter show a remarkable proportionality. Fig.~\ref{fig:structurvsmagnetism} shows also that the low temperature saturation value of the weighted \mb hyperfine field decreases continuously with increasing Co/K-substitution: $B_{\text{hf}}$ at T=4.2 K is reduced from 5.5 T in {\BaFA}~\cite{RotterPRB2008} to 3.0~T, 2.7~T, 1.4~T and 0.9~T for $x/2\approx y = 0.07$, 0.10, 0.13 and 0.16 respectively. Likewise, $T_N$ is reduced to 80~K, 70~K, 40~K and 23~K. Note that plotting $B_{\text{hf}}(4.2K)$ as a function of $T_N$ yields a straight line that passes through the origin as shown in Fig.~\ref{fig:bhf-vs-tn-graph}. From this we conclude, that the underlying mechanism for the magnetic ordering is identically for all compositions.
%
% moessbauer isomer shift comment:
Interestingly in that context is that the {\mb} isomer shifts (IS) at 4.2~K are slightly higher in the purely magnetic samples ($\text{IS}\geq 0.52$~mm/s for {\it x}/2 $\approx$ {\it y} = 0.07 and 0.10) than in the superconducting samples ($\text{IS}\leq 0.51$~mm/s for {\it x}/2 $\approx$ {\it y} = 0.13 and 0.16), see Tab.~\ref{tab:moessparameter}. This means that the local electron density at the iron site is sligthly enhanced in the superconducting samples.

%
%
%Zero Field Muon Spin Spectroscopy
\begin{figure*}
\centering
\includegraphics[width=0.48\textwidth]{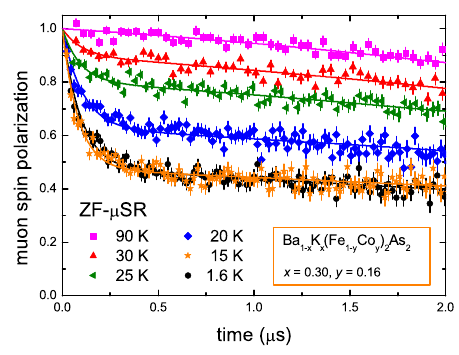} \hfill
\includegraphics[width=0.48\textwidth]{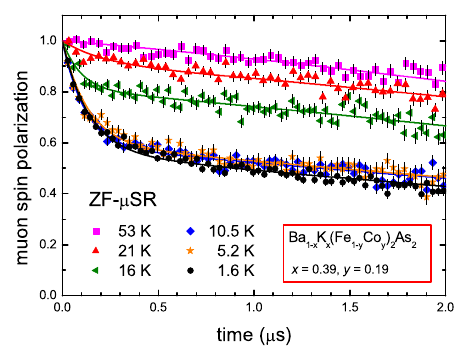} \\
\includegraphics[width=0.47\textwidth]{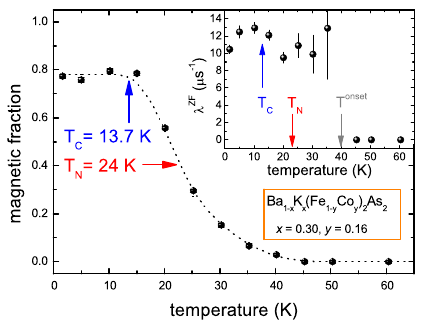} \hfill
\includegraphics[width=0.47\textwidth]{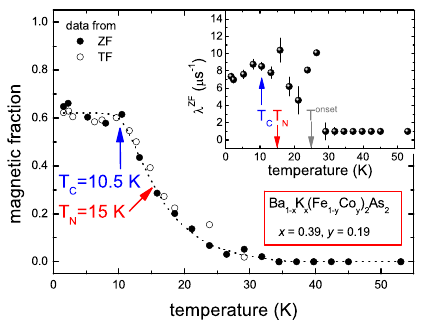}
\caption{Upper panels: Zero field $\mu$SR spectra of {\BaKCo} with {\it x}/2 $\approx$ {\it y} = 0.16 and 0.19. Lower panels: Extracted temperature dependencies of the corresponding magnetic volume fraction and the ZF-$\mu$SR transverse relaxation rate $\lambda^{\text{ZF}}$ (insets). Dotted lines are guides to the eyes.}
\label{Fig:ZF-spectra}
\end{figure*}

In order to verify our results from \mbs and the fitting model, we performed zero field (ZF) \musr measurements ($\mu$SR) on the sample with {\it x}/2 $\approx$ {\it y} = 0.16. It has been shown, that $\mu$SR is an excellent tool to study the volume-selective electronic properties of iron pnictides.\cite{luetkensnaturem2009,marsikprl2010,bernhardprb2012,Wiesenmayer2011,2012arXiv1210.6881} The superconducting properties are directly accessible and $\mu$SR is sensitive to very small ordered magnetic moments. A second sample with {\it x}/2 $\approx$ {\it y} = 0.19 was also studied to confirm magnetism at higher substitution levels. In Fig.~\ref{Fig:ZF-spectra}, zero field $\mu$SR spectra for {\it x}/2 $\approx$ {\it y} = 0.16 (upper panel left) and {\it x}/2 $\approx$ {\it y} = 0.19 (upper panel right) are shown for selected temperatures. At high temperatures, both samples are characterized by a weakly damped Kubo-Toyabe depolarization ($G^{\text{LGKT}}(t,\sigma_{\text{nucl}},\lambda_{\text{nucl}})$) which is typical for the presence of a magnetic field distribution at the muon site due to static nuclear moments only.\cite{hayanobrp1979nuclgausskubotoyabefunction} At temperatures below 40~K ({\it x}/2 $\approx$ {\it y} = 0.16) or 25~K  ({\it x}/2 $\approx$ {\it y} = 0.19) electronic magnetism is evidenced by a strong exponential relaxation of the $\mu$SR time spectra at early times. The absence of an oscillatiory signal is associated to a broad distribution of local fields at the muon site which is typical for chemically substituted pnictides.\cite{marsikprl2010,bernhardprb2012} Accordingly, the time spectra were fitted (solid lines) using the function
\begin{equation}
\begin{array}{rcccl}
	{\text{P}}(t) & = & {\mathcal{M}}               & \times & \left\{2/3\cdot\exp{\left(-\lambda^{\text{ZF}}\cdot t\right)} +1/3\right\}\\[2mm]
	{}            &{+}&  \left(1-\mathcal{M}\right) & \times & G^{\text{LGKT}}\left(t,\sigma_{\text{nucl}},\lambda_{\text{nucl}}\right),
\end{array}
\label{eq:musrzerofield}
\end{equation}
where $\mathcal{M}$ is the magnetic volume fraction and $\lambda^{\text{ZF}}$ the magnetic ZF-$\mu$SR relaxation rate. The nuclear depolarization rates indexed by 'nucl' were fitted for all temperatures simultaneously to avoid a parameter correlation with the magnetic ZF-$\mu$SR relaxation rate.
It is important to note that using ZF-$\mu$SR, the magnetic volume fraction $\mathcal{M}$ can be determined with high accuracy, \emph{viz}. within an error range of a few percent.

The magnetic transition is gradual in temperature as seen by the increase of the magnetic volume fraction in Fig.\ref{Fig:ZF-spectra} (bottom). As an estimate for the N\'{e}el temperature we define --analogue to \mbs -- $T_N$ as the temperature at which 50\% of the sample is magnetically ordered with respect to the low temperature saturation value. We obtain  $T^{50\%}_{\mu}=24$~K for {\it x}/2 $\approx$ {\it y} = 0.16 in agreement with \mbs and $T^{50\%}_{\mu}=15$~K for {\it x}/2 $\approx$ {\it y} = 0.19.

For both samples, the ZF-$\mu$SR transverse relaxation rate $\lambda^{\text{ZF}}$ shows a maximum at ${T_c}$ obtained from ac-susceptibility with a following decrease towards low temperatures. Also at $T_c$, the magnetic volume fraction ceases to increase and remains constant towards lower temperatures. The combination of these two anomalies is reminiscent of the decrease of the magnetic order parameter seen in Ba$_{1-x}$K$_x$Fe$_2$As$_2$\cite{Wiesenmayer2011} and \BaCo\cite{marsikprl2010,bernhardprb2012} at the onset of superconductivity and indicates that magnetism and superconductivity compete for the same electrons.
%
% ---------------------------------------------------------
%
\subsection{Superconductivity}
\begin{figure}
\centering
  \includegraphics[width=0.40\textwidth]{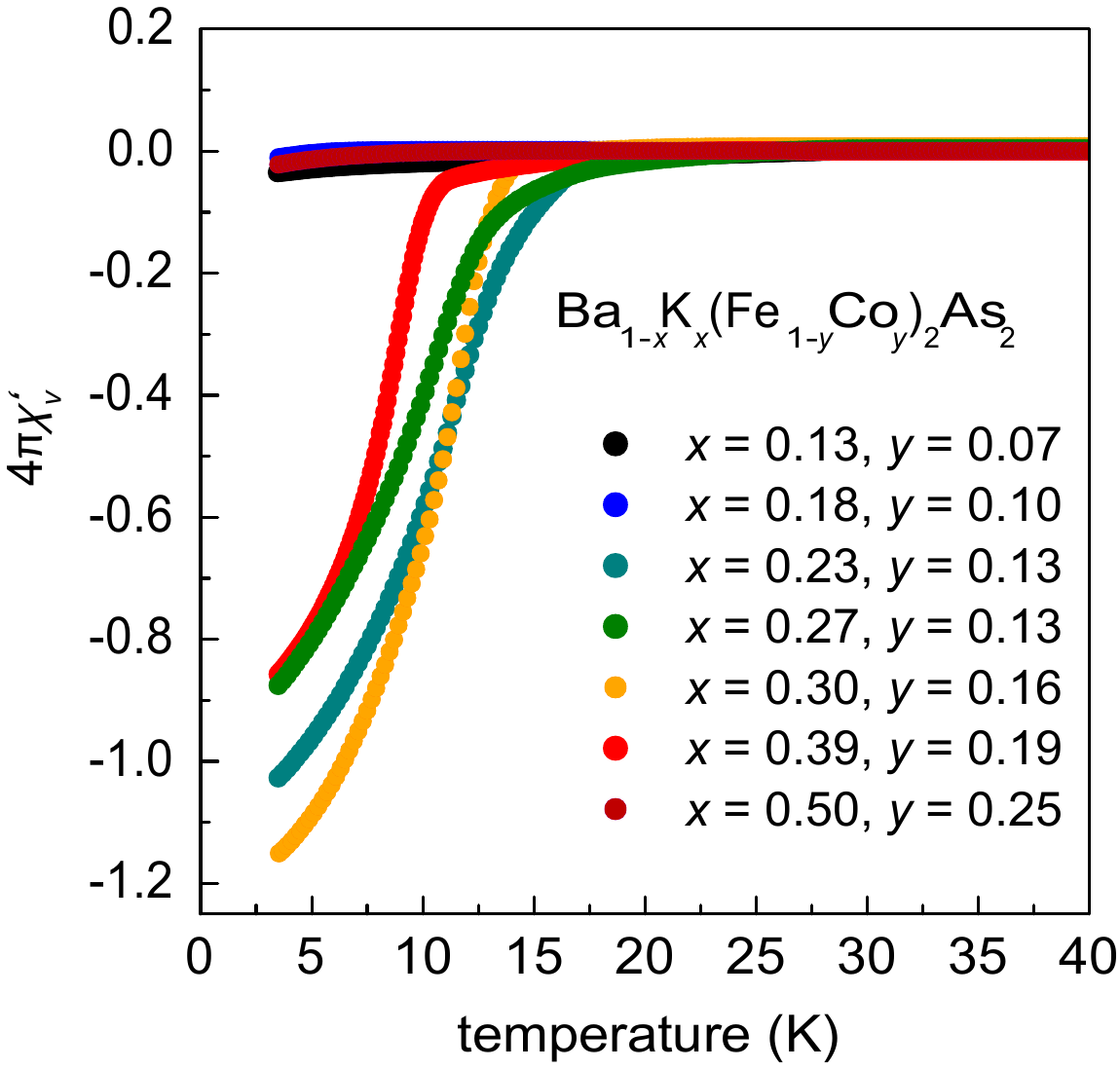}
\caption{Real part of the ac susceptibility signal for charge compensated \BaKCo. For substitution levels $x/2\approx y$ between 0.13 and 0.19, bulk superconductivity is found.}
\label{fig:acs}
\end{figure}

Measurements of the ac susceptibility (see Fig.~\ref{fig:acs}) confirm \emph{bulk} superconductivity for charge compensated \BaKCo with a substitution level ranging from $x/2\approx y=0.13$ to $0.19$. Finally at $x/2\approx y=0.25$, we observe no more diamagnetic response signal, which shows that the superconducting dome spans about 10\% in the cobalt content.

From the low temperature value of $\chi'$, we estimate the superconducting volume fraction $\mathcal{S}$ for a quantitative comparison to the magnetic volume fraction $\mathcal{M}$ extracted from ZF-$\mu$SR: Even if we assume that due to demagnetization corrections $\mathcal{S}$ only becomes 100\% for $4\pi\chi_\nu'=-1.5$, we find a lower boundary for the superconducting volume fraction of $\mathcal{S}=0.9$ for the sample with $x/2\approx y=0.16$ (Fig.~\ref{fig:acs}, yellow curve) and $\mathcal{S}=0.7$ for the sample with $x/2\approx y=0.19$ (Fig.~\ref{fig:acs}, red curve). A comparision with the magnetic volume fractions $\mathcal{M}$ of the identical samples which we precisely determined using ZF-$\mu$SR (see Fig.~\ref{Fig:ZF-spectra}) yields $\mathcal{M}=0.8$,~$\mathcal{S}=0.9$ for $x/2\approx y=0.16$ and $\mathcal{M}=0.6$,~$\mathcal{S}=0.7$ for $x/2\approx y=0.19$. Thus, the overlap of the superconducting and magnetic volume is at least 70\% for a substitution level of $x/2\approx y=0.16$ and 30\% for a substitution level of $x/2\approx y=0.19$. Together with the maximum of the ZF-$\mu$SR transverse relaxation rate $\lambda^{\text{ZF}}$ which we observe at ${T_c}$ in both samples, we conclude from these values that in charge compensated \BaKCo, superconductivity and magnetism coexist \emph{microscopically} at least in a partial volume fraction.

For {\it x}/2 $\approx$ {\it y} = 0.19 we performed transverse field (TF) $\mu$SR~experiments to study the superconducting properties of the non-magnetic volume fraction (1-$\mathcal{M}$) in detail.
In a type-II superconductor, one measures the propability distribution $n(B)$ in the vortex state:\cite{brandtprbrapid1988,SonierRevModPhys.72.769,Sonier.muonTFandDisorder.PhysRevLett.106.127002} Its first moment, the measured averaged magnetic field $\left\langle B\right\rangle$ at the muon site is given by the oscillating frequency in the $\mu$SR time spectrum. Electronic and nuclear dipolar fields as well as the spin polarization of the conduction electrons at the muon site result in a shift of $\left\langle B\right\rangle$ with respect to the applied external field, known as (muon) Knight shift.
For polycrystalline samples, the field distribution from the vortex lattice (VL) can be approximated using a Gaussian function. Therefore, VL formation causes an additional gaussian relaxation rate $\sigma_{\text{SC}}$ in the time domain. From its temperature dependence\footnote{Nuclear and vortex lattice contributions to the measured relaxation rate were separated in the following way: Since the nuclear contribution is a convolution of a lorentzian and gaussian relaxation rate (see Eq.~(\ref{eq:musrzerofield})), the two components were substracted quadratically for the gaussian and linearly for the lorentzian in the frequency (field) domain:
$\sigma_{\text{SC}}(T)=\sqrt{ \sigma^2(T) - \sigma^2_{\text{nucl}} } - \lambda_{\text{nucl}}$. The values for $\sigma_{\text{nucl}}=0.14\,\mu\text{s}^{-1}$ and $\lambda_{\text{nucl}}=0.05\,\mu \text{s}^{-1}$ are obtained from the zero field measurements. } one can estimate the London penetration depth $\lambda$ for T$\rightarrow$0 via the Brandt formula $\sigma_{\text{SC}}^2 = 0.00371 \, \Phi_0^2 \, / \, {\lambda^4}$.\cite{brandtprbrapid1988}

\begin{figure}
	\centering
		\includegraphics[width=0.5\textwidth]{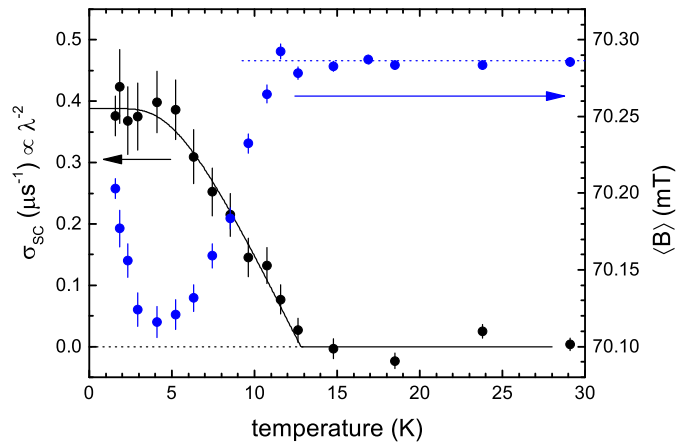}
	\caption{Temperature dependence of the local field $\left\langle B \right\rangle$ (blue dots) and the vortex lattice relaxation rate $\sigma_{\text{SC}}$ (black dots) within the non-magnetic volume fraction of Ba$_{0.62}$K$_{0.38}$Fe$_{1.62}$Co$_{0.38}$As$_2$ ({\it x}/2 $\approx$ {\it y} = 0.19) as determined by a transverse field experiment. The solid black line is a fit within a single s-wave gap scenario (see text for details).}
	\label{fig:y19-700gtf}
\end{figure}

Both measured quantities $\left\langle B\right\rangle$ and $\sigma_{\text{SC}}$ are shown together in Fig.~\ref{fig:y19-700gtf}.
In the non-superconducting (normal) state above $\sim$12~K, $\left\langle B \right\rangle$ remains constant. For polycrystalline iron-pnictides, this value is typically slightly smaller than the applied external field yielding a negative normal-state Knight shift.\cite{luetkens.privcom} The downturn of $\left\langle B \right\rangle$ below $\sim$12~K (Meissner effect) is accompagnied by the appearance of increasing $\sigma_{\text{SC}}$ in the $\mu$SR time spectra and confirm microscopically the superconducting phase transition determined by ac susceptibility measurements (${T_{C,\text{acs}}}=10.5~K$, see Fig.~\ref{fig:acs}).
The orign of the upturn of $\left\langle B \right\rangle$ below $\sim$4~K cannot be conclusively explained. Most likely, it may be associated to a Yosida-like\cite{yosidafunction.PhysRev.110.769} decrease of the spin-susceptibility which is concealed by the diamagnetic shielding %orbital currents
at temperatures between 4 and 12~K.
%for Til: Oberhalb Tc negative Knight Shift dadurch führt singlet-pairing zu negativer "Kopplung" ans externe feld. Die Observable B_local geht HOCH
Also field induced magnetism,\cite{PhysRevLett-103-067010,PhysRevB.82.094512} vortex disorder\cite{Sonier.muonTFandDisorder.PhysRevLett.106.127002} or other effects might cause this anomaly. This topic is worth to study seperately in detail if big enough single crystals of \BaKCo become available.
The temperature depence of $\sigma_{\text{SC}}$ shown in Fig.\ref{fig:y19-700gtf} suggests a nodeless symmetry for the superconducting order parameter. Our data is compatible with a BCS-like single s-wave gap scenario and a corresponding fit\cite{maeterthesis,evtushinskynjp115055069} (solid black line) yields $\lambda(0)=280$~nm and a gap value of $\Delta=1$~meV for $T_c=13$~K.
From these values, we calculate the BCS coupling ratio $2\Delta/k_\text{B} T_c=1.9$ which agrees well to literature data for values of the so-called "smaller gap" observed in various iron pnictide systems.\cite{PhysRevLett-103-067010,evtushinskynjp115055069}
\begin{table*}[p]
\centering
\begin{tabular}{l @{~~~~~} c c c c c c}
\hline\hline
{Sample No.} &  & N27 & N43 & N47 & N45 & N46\\ \hline
{Low temperature state} & afm & afm & afm & {afm+sc} & {afm+sc} & {afm+sc} \\
{K-substitiution} &  $x=0.00$ & $x=0.13$ & $x=0.18$ & $x=0.27$ & $x=0.30$ & $x=0.39$ \\
{Co-substitiution} & $y=0.00$ & $y=0.07$ & $y=0.10$ & $y=0.13$ & $y=0.16$ & $y=0.19$ \\
\hline
{${T_{c,\text{acs}}}  $ [K]} & {0}  & {0} & {0}  & {14}  & {13.7} & {10.5} \\[2mm]
{${T_{S,\text{x-ray}}}$ [K]} & {140} & {90} & {105} & {~50} & {0}   & {0}   \\[2mm]
{${T^{\text{onset}}_{M}}~|~{T^{\text{onset}}_{\mu}}$ [K]} & {155} & {95(5)} & {100(10)} & {50(5)$\,|\,$--}& {29(5)$\,|\,$40} & {--$\,|\,$25} \\[2mm]
%{${T^{sat}_{M}}$ [K]} & {} & {70(3)} & {63(3)} & {28(5))}& {18(5)} & {} \\[3mm]
{$T_N={T^{50\%}_{M}}~|~{T^{50\%}_{\mu}}$ [K]} & {138} & {76(5)} & {70(10)} & {40(5)$\,|\,$--}& {23(5)$\,|\,$24} & {--$\,|\,$15} \\[2mm]
\hline
{$B_{\text{hf}}$(4.2\,K) [T]} & {5.5} & {3.0(1)} & {2.7(1)} & {1.4(1)}& {0.9(1)} & {--} \\[2mm]
%{$B_{\text{hf}}$(4.2\,K)/$T^{onset}_{M}$ [T/K]} & {22} & {32(3)} & {37(3)} & {36(4)}& {33(5)} & {--}  \\[2mm]
{$B_{\text{hf}}$(4.2\,K)/${T^{50\%}_{M}}$ [T/K]} & {0.04} & {0.039(4)} & {0.039(7)} & {0.035(6)}& {0.038(8)} & {--} \\[2mm]% mean=0.036
{IS(4.2\,K) [mm/s]} & {0.53(1)} & {0.531(5)} & {0.526(5)} & {0.509(5)}& {0.504(5)} & {--} \\[2mm]
{$\epsilon$(4.2\,K) [mm/s]} & {-0.04(1)} & {-0.030(10)} & {-0.030(12)} & {-0.034(15)}& {-0.030(15)} & {--} \\[2mm]
%{$\delta$(293\,K) [mm/s]} & {0.407(5)} & {0.412(5)} & {0.407(5)} & {0.392(5)}& {0.390(5)} & {--} \\[3mm]
{$\Gamma/2$(293\,K) [mm/s]} & {0.155} & {0.18(1)} & {0.15(1)} & {0.15(1)}& {0.14(1)} & {--} \\
\hline\hline
%{QS(293\,K) [mm/s]} & {0.11(2)} & {0.13(1)} & {0.09(1)} & {0.10(2)}& {0.12(2)} & {--} \\[2mm]
%{QS($T\geq{T^{\text{onset}}_{M}}$) [mm/s]} & {--} & {0.26(2)} & {0.20(2)} & {0.14(1)}& {0.14(1)} & {--}
\end{tabular}
\caption{Transition temperatures from x-ray diffraction, ac-susceptibility and \musr along with selective hyperfine parameters for {\BaKCo} derived from {\mbs} measurements. The isomer shift IS is given relative to metallic iron; $\epsilon=\left(e^2qQ/2\right)$ is the effective quadrupole shift. Data for \BaFA are taken from Refs.~\cite{RotterPRB2008,Rotter2009}}
\label{tab:moessparameter}
\end{table*}
\subsection{Band Structure Calculations}
To study the development of the electronic and magnetic properties of the \BaKCo compounds, we performed a series of density functional calculations in different approximations. In a simple rigid band picture, a charge compensated substitution ($x/2 = y$) would not yield any changes of the related electronic structure, apart from changes caused by the slight modifications of the crystal structure (see Fig.~\ref{fig:n43str}). % upon substitution.
Therefore, we applied a virtual crystal approximation (VCA) for both the Ba- and the Fe-sites. Since Ba and K act essentially as a charge donor on the $A$-site in the $A$Fe$_2$As$_2$ compounds, and Fe and Co behave rather similar with respect to their bonding behavior, a VCA approach should describe the change of the averaged crystal potential in good approximation. (To simulate the changed electron count due to the Ba-K substitution, only the number of Ba valence electrons was reduced according to the K content, disregarding the different core electrons of both elements.) Since the translation symmetry of the unit cell is preserved, the direct influence of the substitution related scattering on the electronic structure is neglected. This will be part of a future study.

The charge compensated Fe-Co/Ba-K substitution in \BaKCo leads to a fast suppression of the ordered magnetic moment in the stripe antiferromagnetic state with increasing $x/2 = y$ (see upper panel of Fig.~\ref{DOS}, open squares). Our calculations yield a fast, essentially linear decrease of the Fe magnetic moment with increasing Co/K substitution, followed by a sudden drop and the complete breakdown of the stripe AFM order at about $x/2=y\approx0.15$. This result agrees well with the suppression of the measured low temperature saturation value for the weighted \mb hyperfine field $B_{\text{hf}}$ (see Fig.~\ref{DOS}, filled circles). To compare the calculated and experimental Fe moment, $B_{\text{hf}}$ is scaled by a constant factor so that both values match the one of the unsubstituted BaFe$_2$As$_2$ compound.
\begin{figure}
	\centering
	\includegraphics[width=0.45\textwidth]{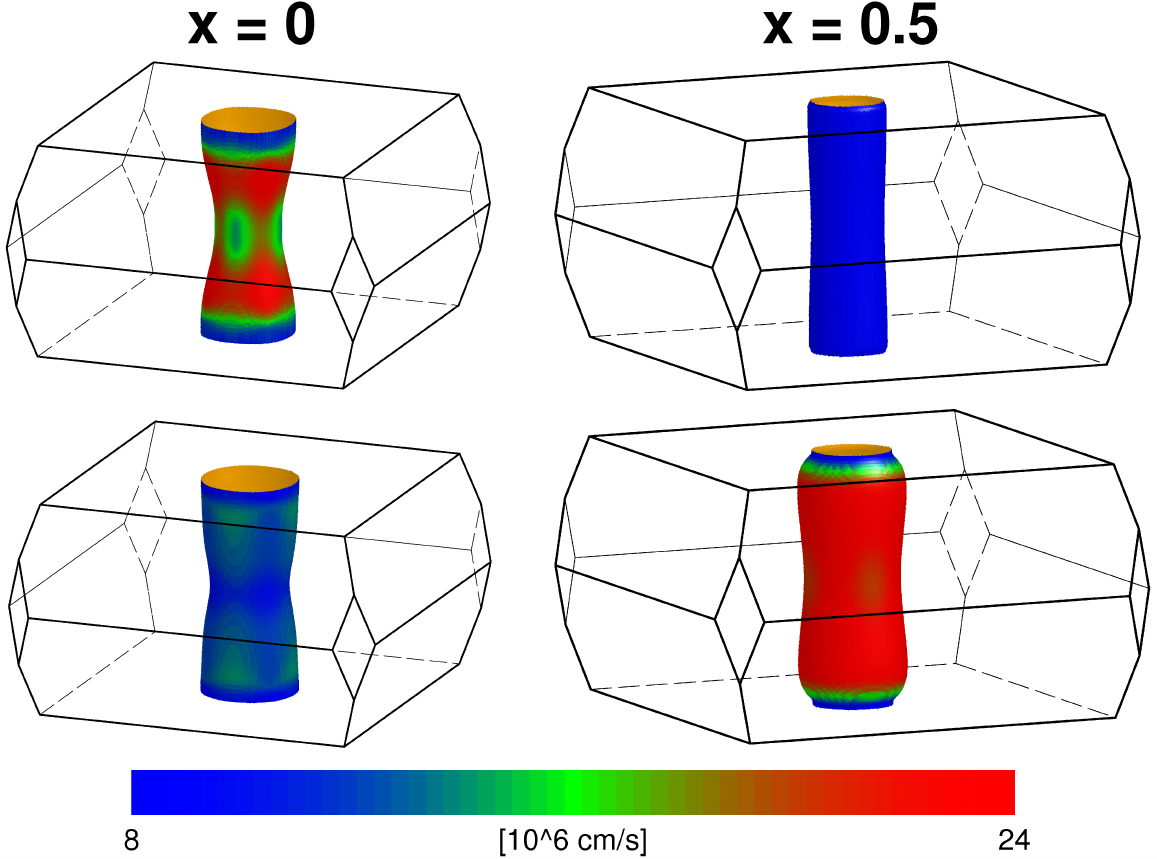}%
	\caption{The calculated hole-related Fermi surface sheets of \BaKCo for $x=y=0$ (left panel) and $x/2=y=0.25$ (right panel). The respective Fermi velocities are indicated by the color-mapping. The shape of the electron surfaces is essentially unchanged in both compounds.}
	\label{FS}
\end{figure}

A closer look onto the related changes of the calculated electronic structure reveals sizable differences in the shape of the Fermi surfaces (see Fig.~\ref{FS}, non-magnetic calculations of the simple tetragonal cell for $x/2=y=0$ and $x/2=y=0.25$), in particular the hole-related sheets. The suppression of the AFM stripe order for $y \ge 0.15$ is likely, at least in part, caused by the reduced nesting of the Fermi surfaces. Another manifestation of the changes in the band structure is shown by the electronic density of states (DOS) $N$ at the Fermi level $\varepsilon_F$: Increasing substitution $y$ results in a sizable reduction of $N(\varepsilon_F)$ which can be assigned predominantly to the Fe 3$d$ states (see Fig.~\ref{DOS}, lower panel). Interestingly, the suppression of the magnetic moment is well compatible with a naive Stoner-like picture. According to the Stoner criterion, which yields a magnetic instability by a divergency of $\chi = \chi_0/(1-I N(\varepsilon_F))$ if $I N(\varepsilon_F) \le 1$, a magnetic ground state would be expected for small $y$: For Fe-3$d$ states with $I \sim 1\pm0.1$, the Stoner criterion is only fulfilled for $N(\varepsilon_F) \ge 1/I$ which is marked by the dashed lines in Fig.~\ref{DOS}. Even if the Stoner criterion is not directly applicable for the formation of the AFM stripe order that involves Fermi surface nesting, it provides at least a qualitative picture for the development of a magnetic instability with respect to Fe-Co/Ba-K substitution and suggests, why the AFM stripe order
is not replaced by a simple FM order when this substitution suppresses the nesting.

In addition to the study of the Fermi surfaces and suppression of the stripe-order magnetism, we tried to separate the influence of the substitution related structural changes (see Fig.~\ref{fig:n43str}) on the electronic structure and the influence of the change of the averaged crystal potential, which was discussed in previous studies.\cite{Suzuki2010} This change originates from the different charge distribution within and between the Fe-As layers caused by the varying (charge compensated) Fe-Co/Ba-K substitution. To this aim, we carried out calculations where (i) only the structure of the parent compound BaFe$_2$As$_2$ was changed simulating the experimentally observed crystal structure (see Fig.~\ref{fig:n43str}), (ii)~substitution only was simulated by the VCA approach keeping the structure of the unsubstituted BaFe$_2$As$_2$, and (iii)~calculations for the "real" compounds combining (i) and (ii). Our calculations indicate that the dominant contribution to the changes of the electronic properties is due to substitution~(ii); the structural changes, in particular the elongation of the $c$ axis, play a minor role. Interestingly, the total changes of the DOS $N(\varepsilon_F)$ in the full calculation~(iii) can be represented as the sum of the changes from (i) and (ii) with high accuracy. This supports the idea that both effects are rather independent and offers the opportunity to fine-tune this type of compound, for instance by applying uniaxial or hydrostatic pressure.
\begin{figure}
	\centering
		\includegraphics[width=0.48\textwidth]{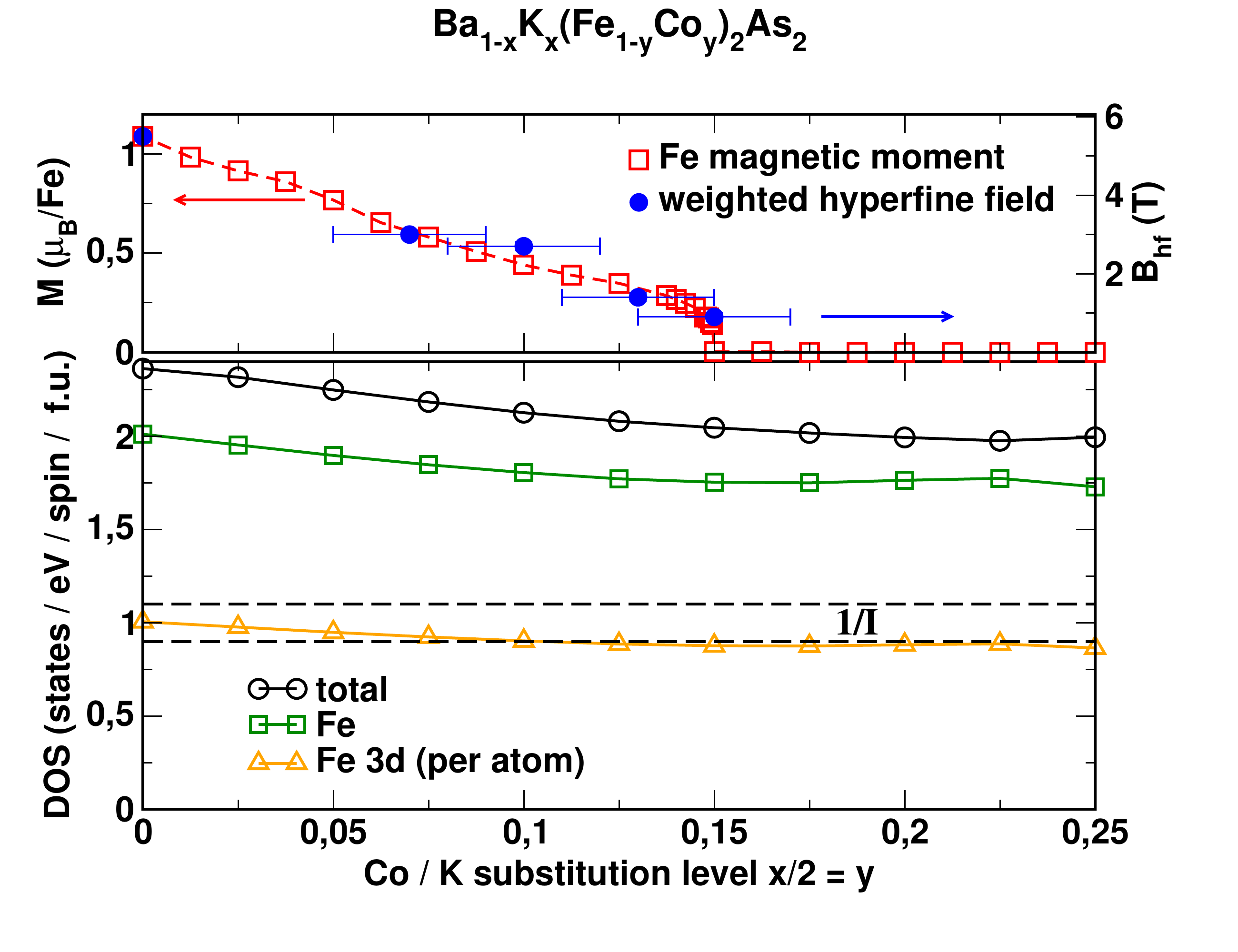}
	\caption{Upper panel: Calculated magnetic momemt (per Fe) in the antiferromagnetic "stripe" phase as a function of the Co / K substitution level $x/2=y$ in \BaKCo together with the weighted hyperfine field $B_{\text{hf}}$(T=4.2\,K). The measured hyperfine fields are scaled by a constant factor to match the calculated Fe-moment for the undoped parent compound \BaFA. Lower panel: Total and partial electronic density of states (DOS) at the Fermi level for \BaKCo for non-magnetic calculations. With increasing substitution level the DOS decreases, and for $y\ge0.11$  the Fe-3d contribution (per Fe atom and spin) drops below the Stoner criterion for a magnetic instability (marked by the dashed lines for $I$(Fe-3d) = $1\pm 0.1$, see text).}
	\label{DOS}
\end{figure}
%
% -----------------------------------------------------------------
% -----------------------------------------------------------------
\section{Discussion}
As we have shown in the previous section, charge compensated \BaKCo with a low substitution level ($x/2 \approx y \leq 0.10$) displays a parent-like antiferromagnetic and orthorhombic state at low temperatures within the whole sample volume. The magnitude of the ordered, magnetic moment and likewise the orthorhombic distortion is subsequently suppressed upon increasing the Co/K substitution level $x/2 \approx y$ (denoted as $y$ in the follwing). Using microscopic techniques, we show that magnetism persists up to $y=0.19$ whereas the orthorhombic distortion is only detected up to $y=0.13$. The onset of magnetic order and structural distortion occur at the same temperature and our data prove a temperature independent, linear relationship between the normalized orthorhombic splitting $\delta_s$ and the weighted hyperfine field $B_{\text{hf}}$, \emph{i.e.} between the magnetic and structural order parameter.
The suppression of the ordered moment is well reproduced by our DFT calculations.
Furthermore, we find bulk superconductivity for $0.13\leq y \leq 0.19$. For the samples with {\it y} = 0.16 and 0.19, we conclude from ZF-$\mu$SR and ac susceptibility data, that magnetism and superconductivity coexist microscopically in a large fraction of the sample volume.
\subsection{Magnetism}
\begin{figure}
	\centering
		\includegraphics[width=0.48\textwidth]{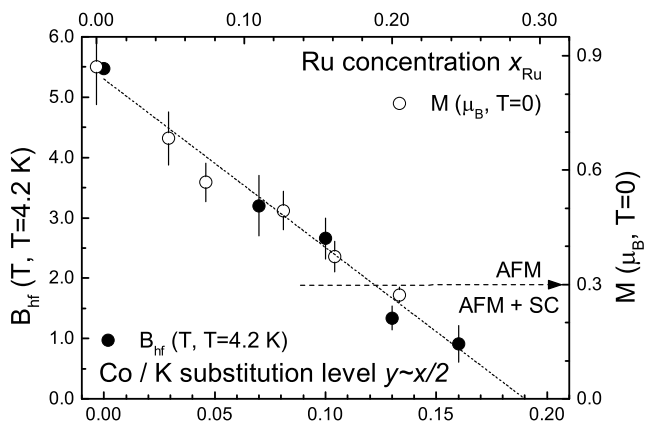}%
	\caption{Ordered magnetic moment at low temperatures in \BaKCo ($B_{\text{hf}}$ for T=4.2\,K from this work, bottom/left axes) and \BaRu (M for T$\rightarrow$0 from Ref.~\cite{Thaler-RuDopedBa122-PhysRevB.82.014534}, top/right axes). Both systems show a linear dependence upon chemical substitution.}
	\label{fig:MagneticMomentComparisiontoRuDoped122}
\end{figure}

For charge compensated \BaKCo, we discuss the supression of magnetism upon increasing the Co/K substitution level by a comparision to the well-studied, nominally isovalent \BaRu system\cite{LaPlace-NMRpercolationlimit-PhysRevB.86.020510,Thaler-RuDopedBa122-PhysRevB.82.014534,kim-PhysRevB.83.054514,Dhaka2011-PhysRevLett.107.267002,kim-PhysRevB.88.014424} because both systems show significant common properties:

The concomitant onset of magnetic order at the structural transition temperature in \BaKCo is reminiscent of the common temperatures for the magnetic and structural phase transition in \BaRu,\cite{kim-PhysRevB.83.054514} and will be discussed seperately in section~\ref{sec:interplaystructuremagnetism}. The suppression of the ordered magnetic moment as a function of chemical substitution in both systems displays a linear behaviour shown in Fig.~\ref{fig:MagneticMomentComparisiontoRuDoped122}.
Both observables for the magnetic order parameter ($B_{\text{hf}}$ for T=4.2\,K from this work and M for T$\rightarrow$0 from Ref.~\cite{Thaler-RuDopedBa122-PhysRevB.82.014534}) are normalized to theirs values in the common pristine system \BaFA and the chemical substitution axes were scaled by a factor of $\sim\,1.5$ in order to get a maximum correspondence. We attribute the fact that less Co/K than Ru substitution is necessary to supress the magnetic moment in the same way to the different sizes of the Ru 4d and the Fe/Co 3d orbitals. Since the Ru 4d orbitals are spatially more extended, the itinerant magnetism is more stable against partial substitution. Note, that a recent XRMS study reports that Ru is spin-polarized with a Ru $L_\text{2}$ edge signal that follows the magnetic ordering of Fe with an in-plane correlation lenght corresponding to more than 700 unit cells.\cite{kim-PhysRevB.88.014424} This may also stabilize the AFM order. In contrast to Ru substitution, Co shows no signatures of spontaneous or induced spin polarization as probed by a combined ${}^{59}$Co and ${}^{75}$As NMR study of Ba(Fe$_{1.8}$Co$_{0.2}$)$_{2}$As$_{2}$.\cite{Ning2008}
%article Ning2008 demonstrates that Co atoms form an alloy with Fe atoms and donate carriers without creating localized moments.
%

Fig.~\ref{fig:MagneticMomentComparisiontoRuDoped122} reveals two further similarities between both systems:
Firstly, superconductivity emerges at a substitution level of $\widetilde{y}\sim0.12$ for \BaKCo and  $\widetilde{x}_{\text{Ru}}\sim0.18$ for \BaRu, respectively, for which the ordered magnetic moment $\mu$ has fallen below a 'critical' value of $\widetilde{\mu} = 0.3\,\mu_B$ (dashed arrow). This will be discussed in section~\ref{sec:discussionsc} below.
Secondly, a linear extrapolation of the ordered macroscopic averaged magnetic moments (dotted lines) yields values for the corresponding substitution levels close to $x_{\text{Ru}}\sim0.3$ and $y\sim0.2$ above which the magnetism is suppressed in both systems. In the \BaRu system, Laplace et al. consider this substitution level (25\% -- 35\%) to be the percolation threshold for magnetic ordering.\cite{LaPlace-NMRpercolationlimit-PhysRevB.86.020510} They also point out, that it is of importance to differenciate between the nominal (macroscopic average) and local (microscopic average) substitution levels $x_{\text{nom}}$ and $x_{\text{loc}}$ depending on how the electronic properties are determined. The Fe layer properties are assumed to be governed by $x_{\text{loc}}$ which may differ from $x_{\text{nom}}$. For \BaRu, they conclude an intrinsically \emph{inhomogenous} electronic state on a local scale which accounts for the coexistence of superconductivity and magnetism. This is in contrast to \BaCo, where \emph{homogeneous} coexistence between superconductivity and magnetism was deduced from the interplay of the superconducting and magnetic order parameter.\cite{Pratt-CoexistenceInCoDoped122-PhysRevLett.103.087001}
For charge compensated \BaKCo with a substitution level of $y=0.16$ and $0.19$, our data shows microscopic coexistence of magnetic and superconducting based on the overlap of the magnetic and superconducting volume fractions at low temperatures. However, the experimental evidence for the interplay of both order parameters \emph{cf.} the reduction of the magnetic ZF-$\mu$SR transverse relaxation rate $\lambda^{\text{ZF}}$ below $T_c$ is weak (see inset of Fig.~\ref{Fig:ZF-spectra}) and has to be studied in more detail. Nevertheless, our results support an homogeneous scenario.

The results of our DFT calculations on \BaKCo quantitatively reproduce the experimentally observed suppression of the ordered magnetic moment upon decreasing $y$ (see Fig.~\ref{DOS}). They also show, that the weakening of magnetism can be directly related to a reduced density of states at the Fermi level consistent to earlier calculations on the endpoint of the $x/2=y$~series (KFeCoAs$_2$, $y=0.5$).\cite{Singh2009} There, the weakening of magnetism was also found as a result of the reduction in the density of states at the Fermi energy N($\varepsilon_{F}$) and this change has been attributed to a shrinking of the in-plane lattice parameter $a$. However, our three-fold DFT study clearly shows, that structural changes are less important for the reduction of N($\varepsilon_{F}$) compared to the effects of chemical (double-)substitution. Furthermore, our results point to a Stoner-like developement of the magnetic instability.
On may understand this in the context of recent results from a valence band photoemission and Auger Electron Spectroscopy study,\cite{PhysRevB.87.134516} showing that Co states have significant Fe 3d character with an almost negligible increased effective onsite Coulomb interaction $U_\text{eff}$. In that work, Kraus et. al. showed also that transition metal states move to higher binding energies with increasing atomic number, contributing less and less to the states close to the Fermi level. From this findings, we deduce that for an in-plane substitution with Co, the density of states N($\varepsilon_F$) is a good quantity for determining the strengh of magnetism in charge compensated \BaKCo.
An effective itinerant scenario for the description of the magnetic instability was also proposed for \BaRu.\cite{Dhaka2011-PhysRevLett.107.267002} In that work, Dhaka et al. suggested, that either magnetic dilution and the associated reduction of an effective Stoner enhancement or the enhancement of impurity scattering leads to the suppression of the magnetic order.
We want to emphasize, that the seemingly contradiction of a random \emph{local} defect beeing responsible for a change in the quantities of an \emph{itinerant} model was reconciled by Berlijn et al.\cite{berlijn-PRL108-207003}% \missing{all: Ich wuerde den folgenden Satz weglassen:} They show that in \BaCo the extra-electron from cobalt takes part of the delocalized electron band and that from within this bath, itinerant carriers conspire to enhance the charge distribution around Co in order to screen the additional proton.
%
%
%
%
% -------------------------------------------------------------------
\begin{figure*}
\centering
{\includegraphics[width=0.48\textwidth]{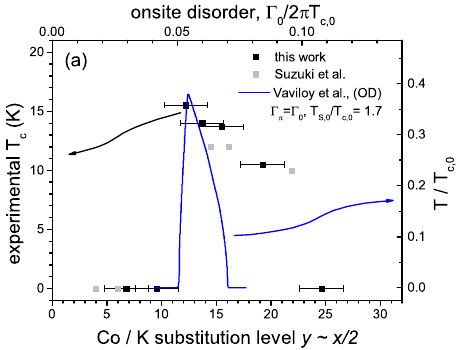}} \hfill
{\includegraphics[width=0.48\textwidth]{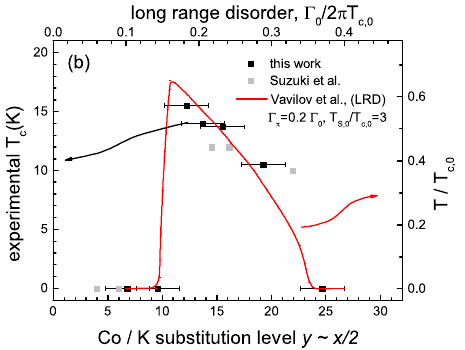}}
\caption{Comparision of experimental ($T_c$ from this work in black and Ref.~\cite{Suzuki2010} in gray) and calculated ($T/T_{c,0}$ from Ref.~\cite{Vavilov2011}) transition temperatures under the assumption that $\Gamma_0/2\pi T_{c,0} \propto n_{\text{imp}} \propto y$. (a) Best possible scaling to a pure \emph{inter}band scattering scenario (OD, blue line). (b) Best possible scaling to a dominantly \emph{intra}band scattering scenario (LRD, red line).}
\label{fig:disordercomparison}
\end{figure*}
\subsection{Superconductivity}
\label{sec:discussionsc}
Our results for the shape and the position of the superconducting dome in charge compensated \BaKCo agree with the results by Suzuki et al.\cite{Suzuki2010} based on resistivity measurements. However, they report a \emph{non-magnetic} superconducting phase for $0.1 \leq y \leq 0.2$ which contradicts our observation of a finite \mb hyperfine field $B_{\text{hf}}$ measured on samples with $y=0.13$ and 0.16 and the magnetic relaxation in the ZF-$\mu$SR time spectra for $y = 0.16$ and $0.19$. Nevertheless, resistivity measurements may not be sensitive to weak magnetic order and therefore, it is possible that superconductivity and magnetism coexist on a microscopic lenght scale.

The onset of superconductivity in charge compensated \BaKCo occurs at a substitution level $y$ between 0.10 and 0.13. As we show in Fig.~\ref{fig:MagneticMomentComparisiontoRuDoped122}, the comparision of the ordered magnetic moment to the \BaRu system yields a substitution level of $\widetilde{y}\sim0.12$ above which the ordered magnetic moment has fallen below a 'critical' value of $\widetilde{\mu}=0.3\,\mu_{B}$\cite{LaPlace-NMRpercolationlimit-PhysRevB.86.020510} giving rise to superconductivity. The value of $\widetilde{\mu}=0.3\,\mu_{B}$ is also found in electron-doped \BaCo for a substitution level of $\widetilde{x}_{\text{Co}}\sim0.05$, above which coexistence and competition between superconductivy and magnetism is reported.\cite{Fernandes-MagPlusSCpairinginElectronDopedBaFe2As2PhysRevB.81.140501}
This is inline with the overall evidence, that for electron doped transition metal-substituted Fe-based 122 superconductors, the substitution level is the salient parameter for the suppression of the magnetic and structural transition to a low-enough temperature, which is a necessary condition for the appearance of superconductivity. However, the span of the superconducting dome is appropriately described using the nominal electron count $e$,\cite{PhysRevB.80.060501,PhysRevB.82.024519} but only as long as a rigid band picture holds and as long as the transition-metal impurity potential shift is small, \emph{viz.} only for cobalt substitution.\cite{PhysRevLett.110.107007}
For systems with effectively isovalent substitution, $e$ has no physical meaning and thus the changes of the electronic properties must be effectively related to the substitution level. We thus conclude, that for charge compensated \BaKCo, the superconducting transition temperature $T_c$ is essentially depending on the substitution level $y$ or in other words, by the density of in-plane impurities.

Vavilov and Chubukov consider the effect of non-magnetic impurities without a change of the density of carriers for iron-pnictides.\cite{Vavilov2011}
They conclude that the spin density wave (SDW) order is supressed stronger than s$^{\pm}$-superconductivity, because intra- \emph{and} interband scattering is destructive for SDW, but only interband scattering is pair-breaking for an s$^{\pm}$-superconductor.
Since the disorder parameter $\Gamma_0/2\pi T_{c,0}$ introduced by Vavilov and Chubukov in their work is proportional to the impurity density $n_{\text{imp}}$, we assume that (i) qualitatively, \BaKCo with {\it x}/2 = {\it y} is an experimental accessible system for this theory and (ii) that quantitatively, $\Gamma_0/2\pi T_{c,0}$ is proportional to the Co/K substitution level $y$. In Fig.~\ref{fig:disordercomparison}, we compare experimental ($T_c$) and calculated ($T/T_{c,0}$) transition temperatures for a pure interband scattering (onsite disorder (OD), $\Gamma_0=\Gamma_\pi$) and a dominantly interband scattering (long range disorder (LRD), $\Gamma_0=5\,\Gamma_\pi$) scenario discussed in Ref.~\cite{Vavilov2011} We find good agreement for the shape of the superconducting dome within the LRD scenario whereas no suitable scaling could be achieved for OD. The value for $T_{S,0}/T_{C,0}=3$ used in the LRD is reasonable for \BaKCo since $T_{S,0}\sim140\,$K in \BaFA and $T_{S,0}\sim40\,$K in \BaK.

Clearly, a picture in which the quasi-particle scattering is due to dominantly \emph{intra}band processes describes our data for charge compensated \BaKCo better. However, both theoretical scenarios predict only a very narrow range $\Delta y$ for a mixed SC and SDW phase (corresponding to $\Delta y < 0.02$) while our experimental findings show coexistence for a quite broad range ($\Delta y \approx 0.10$).
%
%
% ------------------------------------------------------------------
\subsection{Interplay between Magnetism and Structural Distortion}
\label{sec:interplaystructuremagnetism}
\begin{figure}
\centering
  \includegraphics[width=0.48\textwidth]{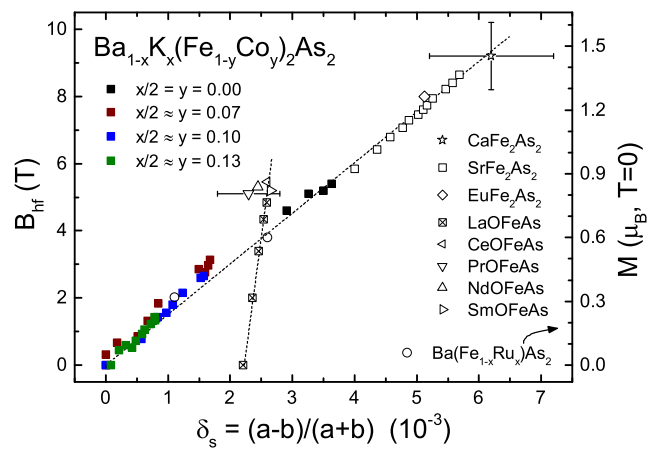}
\caption{Ordered Fe magnetic moment measured by $B_{\text{hf}}$ as function of the orthorhombic splitting parameter $\delta_s$ for \BaKCo with {\it x}/2 $\approx$ {\it y} (same data and colors as in Fig.\ref{fig:structurvsmagnetism}) together with literature data for various nominal charge equivalent 122 and 1111 systems. References are given in Tab.~\ref{tab:structure-hyperfinefield-references}. In the case of \BaRu, the magnetic moment was taken from neutron diffraction data and scaled identically to Fig.\ref{fig:MagneticMomentComparisiontoRuDoped122}. Experimental error bars of our data have been omitted for clarity; error bars are only given for $\text{CaFe}_2\text{As}_2$ and PrOFeAs, where available data differ by more than 10\%. Dotted lines are guides to the eyes.}
\label{fig:implicitplot}
\end{figure}

We now discuss the interplay between magnetism and the orthorhombic distortion in charge compensated \BaKCo. As shown in Fig.~\ref{fig:structurvsmagnetism}, we observe no difference between the structural transition and the onset of magnetic order within experimental resolution. Although the 50\% criterion in our definition of $T_N$ leads to a separation of the magnetic and structural phase transition temperatures, our x-ray and \mb data clearly show emerging magnetism directly below the structural phase transition, similar to the pristine system \BaFA.\cite{RotterPRB2008,Kim-BaFe2Fe2-PhysRevB.83.134522} In contrast, Suzuki et. al. reported a separation of both transitions based on macroscopic resistivity measurements.\cite{Suzuki2010}
The closeness of the structural and the magnetic transition in our samples is better seen if one plots the weighted \mb hyperfine field $B_{\text{hf}}$ as a function of the orthorhombic splitting $\delta_S$, as shown in Fig.~\ref{fig:implicitplot}.

A linear relationship which extrapolates to zero between both quantities becomes evident for charge compensated \BaKCo. A corresponding fit yields $1.5\,$T per $10^{-3}$ for the slope and a negligible value ($<10^{-4}$) for the intercept. This implies that $T_N \approx T_S$ which is a common property of the 122 parent systems \BaFA, $\text{SrFe}_2\text{As}_2$, $\text{EuFe}_2\text{As}_2$, $\text{CaFe}_2\text{As}_2$ and also nominally charge isovalent \BaRu. Furthermore, Fig.~\ref{fig:implicitplot} shows that the low temperature values for the orthorhombic splitting and the ordered magnetic moment of all these systems lie on the same line. In case of magnetic rare earths, values above the rare earth ordering temperature were taken to avoid effects caused by the interplay of rare earth and iron magnetism.\cite{maeterPRB2009} Note that for the 1111 systems where $T_S>T_N$, all mentioned compounds have comparable low temperature values for the structural distortion and ordered iron moment, in stark contrast to the 122 systems where $T_S\approx T_N$. References for the literature data used in Fig.~\ref{fig:implicitplot} is given in Tab.~\ref{tab:structure-hyperfinefield-references}. We also added temperature dependent data for \BaFA and $\text{SrFe}_2\text{As}_2$ to emphasize the linear relationship between the magnetic and structural order parameter we find in \BaKCo.

The implications of the proportionality of the structural and magnetic order parameter in the iron pnictides were firstly adressed by Wilson et al.\cite{WilsonPRB2010} Their analysis of the critical behaviour shows 2D Ising magnetism in $T_N$=$T_S$-materials which transition to a 3D Ising-like character once the structural and magnetic transition are decoupled from each other.
Cano et al.\cite{cano-PhysRevB.82.020408} included harmonic magneto-elastic coupling in the free energy of a Ginzburg-Landau approach and concluded that the difference between the structural and the magnetic transition temperature is crucial for the character of the corresponding phase transition. If $T_S$ is much larger than $T_N$, as it is the case in undoped 1111 systems (illustrated by the finite intercept for LaOFeAs in Fig.~\ref{fig:implicitplot}), both transitions are of second order. If both transition temperatures become closer to each other, the magnetic part of the split transition eventually becomes first-order.

In charge compensated \BaKCo (see Fig.~\ref{fig:structurvsmagnetism}) as well as in \BaRu (Fig.~3 and 4 in Ref.~\cite{kim-PhysRevB.83.054514}) at an intermediate K/Co substitution level or Ru concentration respectively, both transitions are more gradual compared to the step-like transition in the common parent compond \BaFA in which a second-order structural transition is followed by a first-order magnetic transition with $T_S>T_N$.\cite{Kim-BaFe2Fe2-PhysRevB.83.134522} For the case of Ru substitution, the structural and the magnetic transition were assigned to a simultaneous 2nd order transition.\cite{kim-PhysRevB.83.054514} Therefore, charge compensated \BaKCo is likely to behave similar in that sense.
However, the \emph{linear} coupling between the structural and the magnetic order parameter in the vicinity of the phase transitions in \BaKCo from Fig.~\ref{fig:implicitplot} is not compatible with a second order, harmonic magneto-elastic coupling scenario for which a \emph{quadratic} coupling is mandatory by symmetry. Nevertheless, following Refs.~\cite{Kim-BaFe2Fe2-PhysRevB.83.134522,fernandez-nematics-PhysRevLett.105.157003,fernandez-preemptiveNematicOrder-PhysRevB.85.024534} \emph{bilinear} coupling exists between the nematic and structural order parameter which implies a simultaneous nematic-magnetic and structural transition.
Identifying ${T^{\text{onset}}_{N}}$ as the nematic phase transition temperature would at least give qualitatively a consistent explanation for our experimental findings. This includes our observation, that static magnetism coexists together with superconductivity in a tetragonal structure for $0.15 \leq x/2\approx y\leq0.19$.
\begin{table}
\centering
\begin{tabular}{r @{~~~} r@{.}l @{~~~~~} r@{.}l  c }
\hline\hline
 {}     & \multicolumn{2}{l}{$\delta_S$~($10^{-3}$)} & \multicolumn{2}{l}{$B_{\text{hf}}\,$(T)} & \multicolumn{1}{r}{M($\mu_{\text{B}}$, T=0)} \\ \hline
$\text{BaFe}_2\text{As}_2$: & 3&6\cite{Rotter2009} & 5&4\cite{Rotter2009} & {}\\
$\text{EuFe}_2\text{As}_2$: & 5&1\cite{tegel2008EuFe2As2,Mishra_arXiv1304.0595} & 8&0\cite{blachowski-AFe2As2-PhysRevB-83-134410} & {}\\
$\text{SrFe}_2\text{As}_2$: & 5&7\cite{jeschePRB2008} & 8&8\cite{jeschePRB2008,liHFImoessonsrfe2as2} & {}\\
$\text{CaFe}_2\text{As}_2$: & 5&1\cite{CaFe2As2-structure-PhysRevB.78.100506} & 8&2\cite{CaFe2As2-zhiweili} & {} \\
{}   & 6&2\cite{Mishra_arXiv1304.0595} & 8&5\cite{blachowski-AFe2As2-PhysRevB-83-134410} & {}\\
{}   & 6&6\cite{Ni-CaFe2As2-structure-PhysRevB.78.014523} & 10&2\cite{CaFe2As2alzamora} & {} \\
{}   & 7&0\cite{Goldman-CaFe2As2-structure-PhysRevB.78.100506} & 10&2\cite{CaFe2As2PhysRevB79012504} & {} \\
used & 6&2$\pm 1$ & 9&2$\pm 1$ & {} \\
\hline%
LaOFeAs: & 2&6\cite{Nomura-strcture-LaOFeAs-0953-2048-21-12-125028} & 4&9\cite{LaOFeAs-Klauss-PhysRevLett.101.077005} & {} \\
CeOFeAs: & 2&6\cite{maeterLn1111-arXiv1210.6959}  & 5&4\cite{McGuire-NJP-11-025011,maeterLn1111-arXiv1210.6959} & {} \\
NdOFeAs: & 2&4\cite{tian-NdOFeAs-neutrons-PhysRevB.82.060514} & 5&3\cite{McGuire-NJP-11-025011} & {}\\
SmOFeAs: & 2&7\cite{maeterLn1111-arXiv1210.6959} & 5&2\cite{maeterLn1111-arXiv1210.6959} & {} \\
PrOFeAs: & 1&9\cite{luetkens.privcom} & 5&0\cite{McGuire-NJP-11-025011,luetkens.privcom} & {} \\
{} & 2&8\cite{luetkens.privcom} & \multicolumn{2}{c}{} & {} \\
used & 2&3$\pm 0.5$ & 5&0 & {} \\
\hline
\BaRu & \multicolumn{2}{c}{} & \multicolumn{2}{c}{} & {} \\
$x_{\text{Ru}}=0.073:$ & 1&1\cite{kim-PhysRevB.83.054514,kim-PhysRevB.88.014424} & \multicolumn{2}{c}{} & 0.3\cite{kim-PhysRevB.83.054514} \\
$x_{\text{Ru}}=0.205:$ & 2&6\cite{kim-PhysRevB.83.054514} & \multicolumn{2}{c}{} & 0.6\cite{kim-PhysRevB.83.054514} \\
\hline\hline
\end{tabular}
\caption{Low temperature values and references for the orthorhombic splitting and the ordered magnetic moment for selected 122 and 1111 iron-pnictides shown in Fig.~\ref{fig:implicitplot}}
\label{tab:structure-hyperfinefield-references}
\end{table}
%
%
%
%--------------------------------------------------------------------------------
%
\section{Summary}
\begin{figure*}
\centering
{\includegraphics[width=0.4\textwidth]{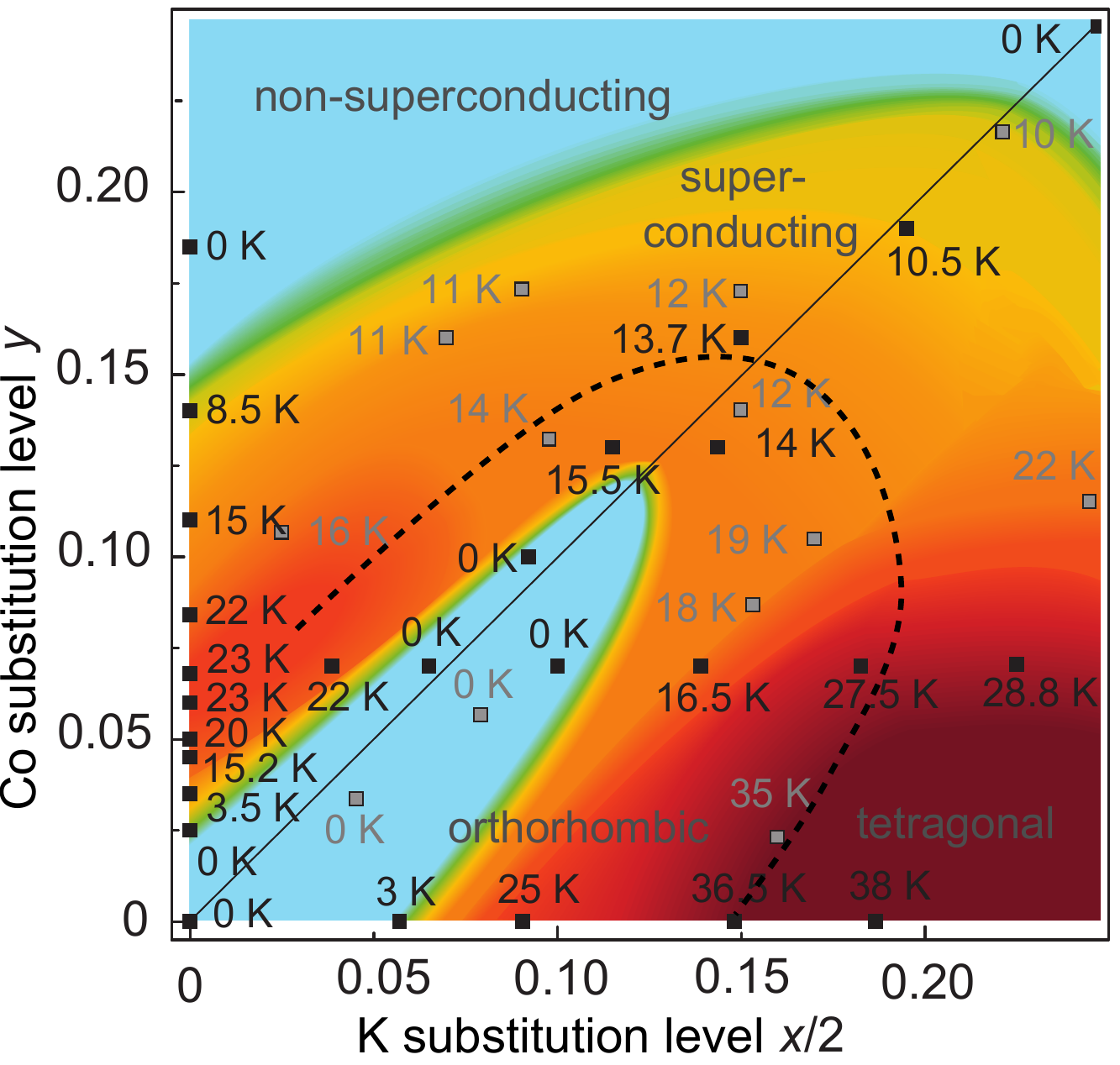}} \hfill
{\includegraphics[width=0.5\textwidth]{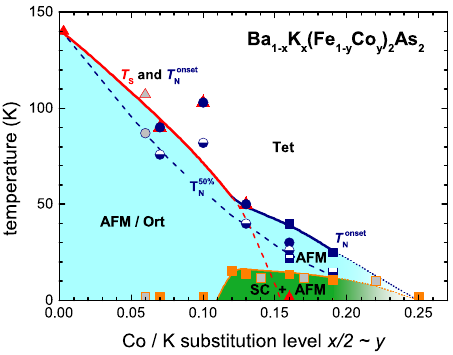}}
\caption{Left panel: Full phase diagram of \BaKCo for $x/2$ and $y\leq$~0.25: Black squares refer to the corresponding superconducting transition temperatures for {\BaK},\cite{Rotter2008} {\BaCo}\cite{Ni2008b} and {\BaKCo} from this work. Grey squares for \BaKCo are taken from Ref.~\cite{Suzuki2010} A non-superconducting groundstate is depicted in blue and superconductivity is shown color-coded from yellow (for $T_c \approx 10\,$K) to dark red ($T_c\approx 40\,$K). Dashed line: Tentative border for the existance of the orthorhombic phase. Right panel: Phase diagram of charge compensated \BaKCo. Data points refer to results from x-ray diffraction (red triangles), \mbs (blue circles) and muon spin relaxation (blue squaeres); $T^{\text{onset}}$ in full symbols, $T^{50\%}$ in half symbols. Data points for the corresponding transition temperatures from Ref.\cite{Suzuki2010} were added in grey.}
\label{fig:phase}
\end{figure*}

The experimental results on charge compensated \BaKCo with $x/2\approx y$ are summarized in a phase diagram in Fig.~\ref{fig:phase}~(right panel) in comparision with full phase diagram of \BaKCo for $x/2\leq0.25$ and $y\leq 0.25$ (left panel). In the charge compensated state, the reported\cite{Suzuki2010} transition from an antiferromagnetic metal ($x/2\approx y\leq0.10$) with an orthorhombic structure to a tetragonal, non-magnetic ground state ($x/2\approx y\geq0.25$) was confirmed. However, local probe techniques evidences that static magnetic order extends to a higher substitution level ($x/2\approx y\leq0.19$) than reported previously ($x/2\approx y\leq0.10$).\cite{Suzuki2010} Moreover, magnetism coexists in a tetragonal structure together with superconductivity for $0.15 \leq x/2\approx y\leq0.19$.

We have discussed the electronic phase diagram of charge compensated \BaKCo within an empirical model which indirectly includes the local defects in an effective Stoner-like picture.
The shape of the superconducting dome can by explained by the introduction of disorder due to non-magnetic impurities ($n_{\text{imp}}$) to a system with a constant charge carrier density.\cite{Vavilov2011} A comparision of the $T_c$-$n_{\text{imp}}$ phase diagrams suggests that the quasi-particle scattering in charge compensated \BaKCo is dominantly due to intraband processes. The measured temperature dependence of the magnetic penetration depth for the sample with $x/2\approx y = 0.19$ suggests a nodeless symmetry for the superconducting order parameter in accordance with the predicted $s^{\pm}$ pairing symmetry. Superconductivity in effectively carrier-free \BaKCo is weaker ($T_\text{c,max}\sim15\,K$) compared to carrier doped \BaK (38 K) and \BaCo (23 K) since it is allways competing with magnetism.

Furthermore, our results prove that magentism is present throughout the whole superconducting dome.
From this finding in addition with the temperature independent, linear relationship between
the magnetic and structural order parameter defined as the normalized orthorhombic splitting $\delta_s$ and the weighted hyperfine field $B_{\text{hf}}$ (see Fig.\ref{fig:implicitplot}), we conclude that for effectively charge compensated \BaKCo, the orthorhombic lattice distortion and the onset of superconductivity is controlled by the magnetic, potentially nematic-magnetic, instability.
\begin{acknowledgments}
This work was financially supported by the German Research Foundation (DFG) within the priority program SPP~1458 projects JO257/6-1 and KL1086/10-1. Additional DFG support within the framework of the Research Training Group GRK~1621 is acknowlegded. Part of this work was performed at the Swiss Muon Source (Villigen, Switzerland). T.G. thanks Johannes Spehling and Rajib Sarkar for valuable discussions.
\end{acknowledgments}
\bibliography{references}

\end{document}